# *Colloquium*: Artificial spin ice: Designing and imaging magnetic frustration

*Cristiano Nisoli*
*Theoretical Division,*
*Los Alamos National Laboratory,*
*Los Alamos, New Mexico, 87545, USA.*

*Roderich Moessner*
*Max Planck Institute for the Physics of Complex Systems, 01187 Dresden, Germany.*

*Peter Schiffer*
*Department of Physics and the Frederick Seitz Material Research Laboratory,*
*University of Illinois at Urbana-Champaign, Urbana, Illinois 61801, USA.*

**Abstract**

Frustration – the presence of competing interactions – is ubiquitous in the physical sciences and is a source of degeneracy and disorder, which in turn give rise to new and interesting physical phenomena. Perhaps nowhere does it occur more simply than in correlated spin systems, where it has been studied in the most detail. In disordered magnetic materials, frustration leads to spin-glass phenomena, with analogies to the behavior of structural glasses and neural networks. In structurally ordered magnetic materials, it has also been the topic of extensive theoretical and experimental studies over the past two decades. Such geometrical frustration has opened a window on a wide range of fundamentally new exotic behavior. This includes spin liquids in which the spins continue to fluctuate down to the lowest temperatures; and spin ice, which appears to retain macroscopic entropy even in the low temperature limit where it enters a topological Coulomb phase. In the past seven years a new perspective has opened in the study of frustration through the creation of *artificial frustrated magnetic systems*. These materials consist of arrays of lithographically fabricated single-domain ferromagnetic nanostructures that behave like giant Ising spins. The nanostructures' interactions can be controlled through appropriate choices of their geometric properties and arrangement on a (frustrated) lattice. The degrees of freedom of the material can not only be directly tuned, but also individually observed. Experimental studies have unearthed intriguing connections to the out-of-equilibrium physics of disordered systems and non-thermal 'granular' materials, while revealing strong analogies to (spin) ice materials and their fractionalized magnetic monopole excitations, lending the enterprise a distinctly interdisciplinary flavor. The experimental results have also been closely coupled to theoretical and computational analyses, facilitated by connections to classic models of frustrated magnetism, whose hitherto unobserved aspects have here found an experimental realization. We review the considerable experimental and theoretical progress in this field, including connections to other frustrated phenomena, and we outline future vistas for progress in this rapidly expanding field.

## I.    Introduction

In many-body physics, complex behavior can arise in the form of collective phenomena even from simple interactions between elementary building blocks; superconductivity and the fractional quantum Hall effect, involving Cooper pairs and fractionally charged Laughlin quasiparticles respectively, are prime examples. In fact, the effort to understand natural materials in terms of new phases and their excitations emerging at low temperature has dominated condensed matter physics research over the last decades.

Recently, the study of collective phenomena has begun to exploit tailor-designed structures of desired properties, which can also be directly visualized. In the case of magnetic materials, this new approach uses advances from the nanosciences—in lithography, atomic-scale microscopy and thin film growth—to customize nanoscopic magnetic degrees of freedom and their geometric arrangement, even allowing the read-out of their states microscopically in real time. This field has taken on the name of artificial frustrated magnetism, on account of the initial focus on the physics of systems with competing interactions. It has opened a new window on collective behavior, manifested in thermodynamic and dynamical properties, both in and particularly out of equilibrium.

"Artificial spin ice" (Wang, 2006) is a two dimensional array of magnetically interacting nanoislands or nanowire links whose magnetic degrees of freedom can be visualized directly in real space through a variety of  techniques (*e.g.*, magnetic force microscopy, Lorentz microscopy, neutron scattering, Hall effect). Introduced initially to mimic the frustrated behavior of naturally occurring spin ice materials and to reproduce celebrated two-dimensional models of statistical mechanics, it has raised many distinct issues and developed into its own field of study. Understanding these systems has turned out to require a novel combination of concepts from fields as diverse as classical correlated spin models, disordered systems, information theory, granular media and micromagnetics, with the ground state entropy of ice and the dynamics of magnetic monopole excitations even putting in prominent appearances. Reflecting its multidisciplinary nature, it has attracted physicists from a broad range of backgrounds, and work in the field has evolved with close connections between experiments and theory.

This Colloquium provides a broad introduction to this young and rapidly developing area of research. We summarize key experimental and theoretical developments of the last seven years, along with the conceptual groundwork that has been laid. We do not aim at an exhaustive survey of the literature; rather our goal is to provide a solid point of reference to scientists interested in learning about the field – establishing a common introduction to, and for, the heterogeneous artificial frustration community.

This article is organized as follows. We first provide an overview of the central themes whose interplay gives rise to new phenomena: tunable degrees of freedom and their geometric arrangement, frustration and the resulting generation of low-energy scales and emergent phenomena, the effect of local constraints on dynamics and equilibration, and finally the role of disorder.

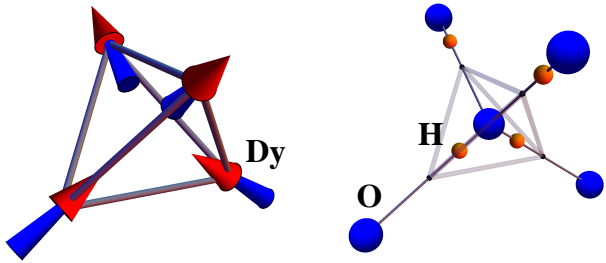

Figure 1. (Color Online). The Ice Rule: In (cubic) water ice (right) $I_c$, oxygen ions are arranged at the sites of a diamond lattice and are hydrogen-bonded by an intermediate proton (red). The lowest energy configuration consists of two protons close to each oxygen atom and two further away. Such ice states map onto spin configurations (left) in which a spin pointing toward the oxygen atom represents a close hydrogen atom. Therefore in the ground state the sum of all the spins in a vertex should be zero: two spins pointing out, two pointing into a tetrahedron (from Castelnovo 2012). The spin model captures the behavior of the spin ice materials, in which ferromagnetically interacting moments are constrained to point directly into or out of the tetrahedra.

The remaining material fleshes out this picture by providing the requisite details. Through an account of the experimental progress that has brought the field into existence, we lay the groundwork by introducing the model systems and probes to study them. We then discuss central theoretical and experimental aspects of artificial frustrated magnets. We conclude by examining a range of offshoots of artificial frustrated magnet research as well as considering possible future directions for the field.

## II.    Context

The ability to control degrees of freedom and manipulate their interactions underpins much of modern applied science. Advances in the field of semiconductors and lithography now allow us to nanostructure quasi-two-dimensional materials with few constraints. In artificial spin ice, such methods are used to deposit single-domain magnets of submicron dimension into two-dimensional arrays. By appropriately designing the shape, size and composition of the magnetic structures, we can control properties such as individual magnetic moments, anisotropy, and coercive field, and therefore tune interactions and responses to an external field. The geometrical arrangement of multiple such structures in close proximity is the essential step underpinning the collective behavior, on which we will focus.   As in conventional magnetism, the geometry of the moment arrangements in a lattice – square, triangular, honeycomb or kagome, for instance – has tremendous ramifications, and *here, it can be chosen at will*. Particular emphasis has been placed on *frustrated arrangements*, whose properties have long been of interest.

### II.a  Frustration in water Ice and spin Ice

Frustration in a physical system emerges from the impossibility of simultaneously minimizing all interactions. It can arise from intrinsic structural disorder, as in spin glasses, or in a regular geometry that carefully balances competing interactions. Pioneering works on geometrical frustration date back to the 1920's, when Pauling explained Giauque's measurements of the zero temperature entropy of water in terms of multiple choices in allocating hydrogen bonds between $H_2O$ molecules in ice (Giauque, 1933; Pauling, 1935). A given oxygen atom in water ice is situated at a vertex of a diamond lattice and has four nearest neighbor oxygen atoms, each connected via an intermediate proton (Figure 1). The proton is not centered between the two surrounding oxygens, but rather is positioned closer to one or the other. The lowest energy state has two protons positioned close to the oxygen and two protons positioned further away, forming a "two-in—two-out" state.  Such states are said to obey an *ice rule*, which can be mapped to a spin model possessing an extensive degeneracy of states (Anderson 1956). In the 1960's and 1970's, a two dimensional analogue of the ice system (the six-vertex model) and its many generalizations were studied extensively and often solved exactly. (Lieb, 1967; Lieb 1971; Wu 1969; Baxter, 1982).

Experimental studies of frustration received a further boost in the early 1990's. Certain magnetic materials displayed unusual behavior characterized by lack of conventional magnetic ordering down to very low temperatures, sometimes orders of magnitude below the energy scale of the interactions among the magnetic moments, as a consequence of geometrical frustration; for reviews from differing perspectives, see (Ramirez 1994; Lacroix 2011; Moessner 2006; Balents 2010*)*. Some of these materials exhibit spin-glass-like behavior, such as history and time dependent properties, at the lowest temperatures – even in the presence of apparently only minimal levels of the structural disorder that has traditionally been associated with such glassiness (Mydosh, 1993). Other materials display strong spin fluctuations down to the lowest accessible temperatures, thus suggesting a new sort of "spin liquid" phase. Most of these materials have antiferromagnetic interactions mainly between pairs of neighboring spins, i.e., with an exchange interaction for a pair of spins minimized when the spins are oppositely oriented.

Harris and coworkers (Harris 1997) initiated a study of a group of pyrochlore materials such as $Ho_2Ti_2O_7$ that appeared to break the mold. Unlike other geometrically frustrated magnets, the moments exhibit a net ferromagnetic interaction between nearest neighbor spins – i.e., the spins energetically preferred to have their moments aligned. Such ferromagnetic interactions are usually considered antithetical to frustration, but the low temperature state of the moments was not ordered, indicating that strong frustration did exist in this system. This frustration arises because the magnetic moments reside on a lattice of corner-sharing tetrahedra, and they are constrained to point either directly toward or away from the center of a tetrahedron (and consequently away from or toward the center of the neighboring tetrahedron). Geometrically, this is exactly the same configuration as the spin model of hydrogen ion positions in water (Figure 1): with the ferromagnetic interaction, the low energy state of any tetrahedron obeys the ice rule, with  two spins pointing in and two spins pointing out.  Harris and coworkers noted this similarity, which was confirmed by measurements of the residual state entropy by Ramirez *et al*. (Ramirez 1999).

The spins in these spin ice materials (which include a range of Dy and Ho pyrochlores) have large moments, so that they can often be described classically (Siddharthan 1999). They have emerged as important model systems with exotic field-induced phase transitions and unusual glassiness. The theoretical investigation of this system has also revealed a range of surprises, including the discovery that the effective nearest-

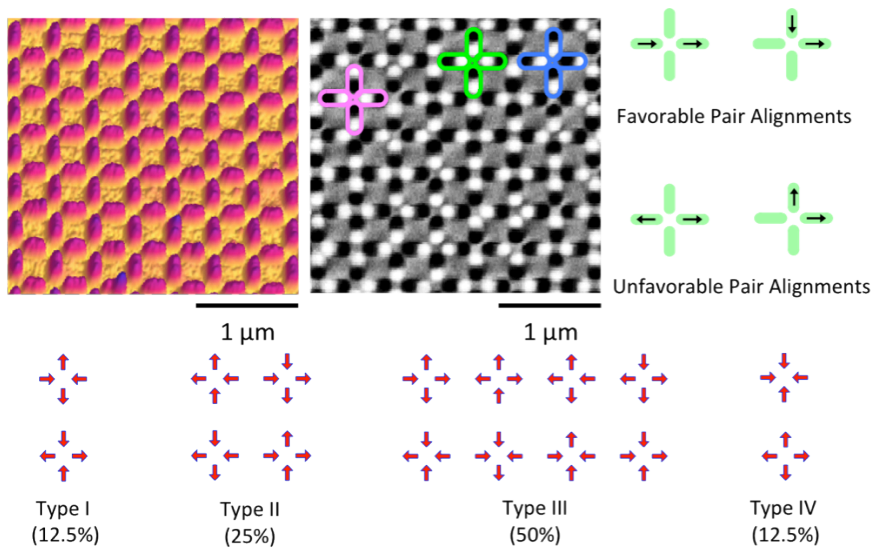

Favorable Pair Alignments

Unfavorable Pair Alignments

1 µm

1 µm

Type I (12.5%)   Type II (25%)   Type III (50%)   Type IV (12.5%)

Figure 2. (Color Online.) Artificial Spin Ice (shown on top left as an AFM image) allows for direct visualization of its magnetic degrees of freedom through MFM (top center) and other techniques (see below in the text). Energetically favorable and unfavorable dipole interactions are described in the top right. Bottom: four legged vertices have $2^4 = 16$ possible moment configurations, which separate into four symmetry-distinct types, shown here with the relative frequency, which corresponds to random assignation of the moments. Note that even in the low energy configurations, some interactions are frustrated, but the lack of degeneracy beside spin inversion in Type I vertices provides an ordered antiferromagnetic ground state.(Wang 2006).

neighbor ferromagnetic interaction arises in part from dipolar coupling among the spins in the system (Siddharthan 1999; Bramwell 2001). Perhaps most excitingly, the fundamental spin excitations in these materials behave like magnetic monopoles (Castelnovo 2008), linked by effective "Dirac strings" of reversed magnetization and exhibiting a magnetic Coulomb interaction. The inverted commas around Dirac strings are in order here, as the reversed magnetisation is observable, and hence these strings, unlike those envisioned by Dirac, do not e.g. impose any conditions on the quantization of electric charge. These excitations have since been probed experimentally (Bramwell 2009; Fennel 2009; Kadowaki 2009; Morris 2009). For a review, see (Castelnovo 2012).

### II.b The ice model and magnetic monopoles

While clarifying the physics of ice — a substance that our ancestors had reason to think about long before the advent of micromagnetics — Lieb's solution of the ice model in two dimensions was an important milestone in the study of critical phenomena. It provided a concrete example of a two-dimensional phase transition not belonging to the Ising universality class (Lieb 1967). Most importantly in the present context, the ice models present the simplest setting in which phenomena of great conceptual importance, such as residual entropies and emergent gauge structures, appear in transparent yet non-trivial ways.

Ice models involve Ising spins residing on the bonds of a lattice. They are also called vertex models because an assignment of energy is given to configurations of spins converging in a vertex. For instance, in the square lattice, the ice rule specifies that half the arrows point into, and the other half out of, each vertex, and the model of configurations satisfying this constraint is called the six-vertex model (Baxter 1982). If one considers the spins as dumbbells of magnetic charge, then the ice rule guarantees charge neutrality in each vertex, and excitations above the ice-rule manifold correspond to magnetic monopoles (a subject that is expanded upon below).

Different generalizations of the ice rule have been proposed in lattices with odd coordination number z, before artificial spin ice came into existence. There, it is not possible to have local charge neutrality, as a sum of an odd number of +1's and -1's cannot add up to zero. Now, one generalization is to consider states with a maximally balanced number of bonds pointing in and out (Wills 2002): each vertex must have *minimal* charge, either +1 or -1. Another one, applicable to bipartite lattices only, is more stringent: the demand is for one sublattice to have charge +1, and the other sublattice -1 throughout (Udagawa 2002, Moessner 2003). We shall see below that these rules define different phases (Ice I and Ice II) of the so-called honeycomb artificial spin ice.

### III.   Artificial spin ice: Basic Structures

While research on geometrically frustrated magnetic materials was developing rapidly, the spin ice systems with their large spins and almost classical behavior suggested a new approach to the study of geometrical frustration, based on the lithographic creation of magnetic nanoislands. Such islands, which are usually a few tens of nanometers thick and have lateral dimensions of the order of 100 nm, are typically fabricated using electron beam lithography. The study of magnetic properties of single nanoislands had already reached a considerable level of maturity both experimentally and theoretically (Bader 2006).

In such small structures, the shape anisotropy, in which a separation of scales in the demagnetizing field in different directions creates an easy axis, can determine the size and moment direction of the ferromagnetic domains, i.e., of the regions in which the atomic moments are all aligned along a particular direction (Imre 2006). For example, in a sufficiently small, elongated ellipsoidal island or a nanowire link in a network, the magnetostatic energy is minimized when the moments align along the long axis. Such islands or links therefore effectively behave as large spins, parameterized by the Ising variable $S = \pm 1$, which encodes which direction they point along the long axis.

In the particular case of single-domain ferromagnetic islands, the islands can be placed in close proximity to each other and arranged geometrically in virtually any two-dimensional configuration. The magnetostatic coupling between neighboring islands can then influence the relative moment directions, and an array of closely spaced islands constitutes an interacting many-body spin system.

The concept of "artificial spin ice" came with the recognition that such a system of ferromagnetic islands could replicate much of the physics of the spin ice systems in a two dimensional model system of ferromagnetic islands. However, these systems proved to be much more complex and revealing than expected.

### III.a Artificial square ice

Wang *et al*. (2006) created such arrays of islands from thin films of permalloy, an iron-nickel alloy with isotropic magnetic properties, in a square geometry shown in Figure 2. If one considers a single vertex of four islands, then the lowest magnetostatic energy states have two moments oriented in toward the center and two oriented away from the center, in direct analogy to the tetrahedra of spin ice materials. For this square artificial spin ice, the intrinsic frustration is similar to that of the two-dimensional square ice model (Baxter 1982), with the important feature that perpendicular islands interact more strongly than parallel ones, so that degeneracy of the ice rule is lifted. This leads to a unique antiferromagnetic ground state, and therefore to an absence of residual entropy. In this sense, when only the interactions at the vertex are considered, artificial square ice is a physical realization of a generalized F-model (Rys 1963, Lieb 1967, Lieb 1971, Wu 1969) rather than of the ice model.

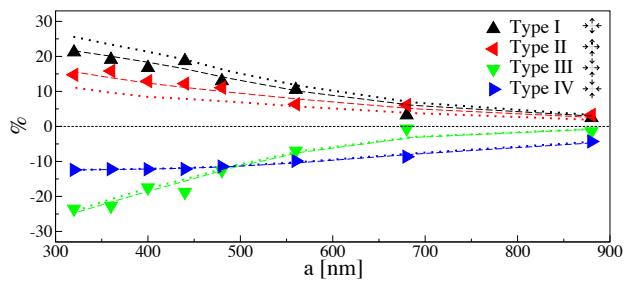

Figure 3. (Color Online.) Suppression of non-ice vertices in artificial spin ice. The excess percentages of different vertex types compared to a random arrangement, plotted as a function of the lattice spacing of the underlying square array, approach zero for the largest lattice spacing, i.e., weakest interactions. Symbols: experimental data from (Wang 2006); lines: theoretical data from dynamical modeling of (Möller 2006).

Wang *et al*. produced square arrays of different lattice constants, all with the same island size, and subjected them to rotational demagnetization—more of this method of preparation later (Wang 2006 and 2007; Ke 2008). They then directly imaged the orientation of magnetic moments via magnetic force microscopy (MFM). By counting the different kind of vertices, they demonstrated a suppression of non-ice vertices and hence an excess of ice vertices (Fig. 3), Wang 2006; Nisoli 2008). An analysis of correlations between spins confirmed that nearest neighbor correlations are dominant and long-range order is absent.

These early experiments on square ice established the possibility of producing two-dimensional magnetic nanoarrays in which the islands interact with sufficient strength to display collective phenomena. Indeed, a simple estimate of the interaction energies (depending or geometry and material of realizations, the moment of each island is approximately $3 \cdot 10^7$ Bohr magnetons, estimated from permalloy, and the separation between islands is of the order of tens of nanometers) yields an energy scale of around $10^4$-$10^5$ K. An analysis of the equilibrium properties of the corresponding effective ice model showed that the experimental system should be well into its ordered state (Möller 2006) at room temperature, at which the experiments were undertaken. This implied that the actual experimental state of the system is athermal because — as we detail below — thermal fluctuations at reasonable working temperatures could not induce spin flips. An explanation of experiment thus needs to take into account the dynamics of the AC demagnetization to which the artificial spin ice was subjected.

Dynamical studies of these systems were initiated by Möller and Moessner, who proposed an approach that assumed islands flip independently, provided that the energy gain from flipping in the applied (rotating) field exceeds a threshold value, as implied by a non-zero coercive field. A second fitting parameter, the attempt rate for island flips, allowed accounting for the experimental data quantitatively (Möller 2006), as depicted in Fig. 3: there, the dotted line displays the best fit for a model of thin dipolar needles. This systematically overestimates the frequencies of Type I vertices, at the expense of those of Type II. By using parameters determined from micromagnetic simulations (Nisoli 2007, Wang 2006) rather than the simple dipolar needle model, the discrepancy with the data disappears as shown by the dashed line in Fig. 3 (Möller 2006). Budrikis and collaborators then pursued this line of modeling to simulate the effects of disorder (Budrikis 2010-2012). In a different approach, Nisoli and coworkers attempted to frame the issue in the context of the statistical mechanics of a granular material (Nisoli 2007, 2010), and found a description in terms of an effective temperature, controlled by the external magnetic drive.

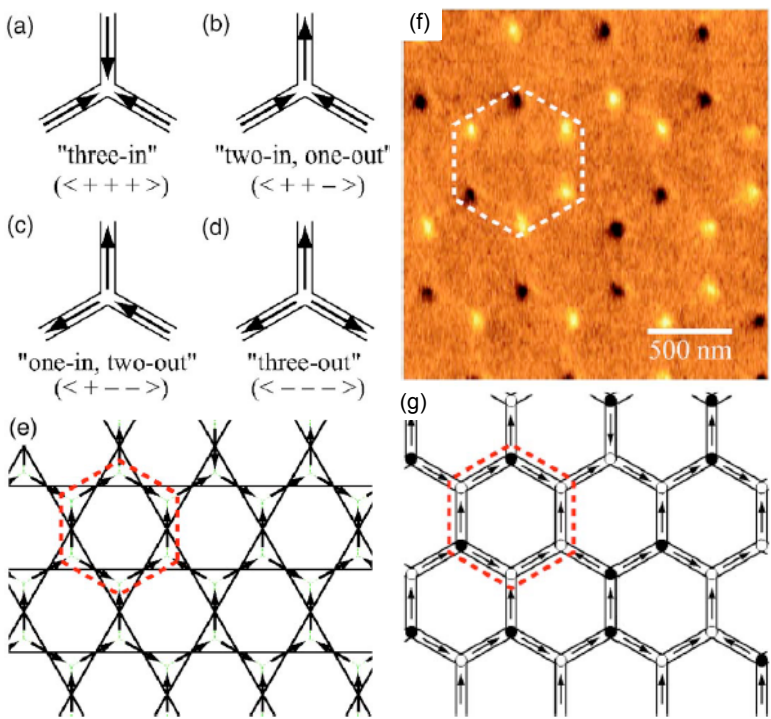

Figure 4. (Color Online.)Honeycomb Ice. (a)–(d) Possible moment configurations at a vertex. (e) The spins in honeycomb ice (red dashed lines) are arranged along the sides of a kagomé lattice (solid line). (f) Schematics of the honeycomb lattice obeying the pseudo-ice rule (two-in/one-out and vice versa). Because of the odd coordination of the lattice, vertices of the honeycomb, which form a hexagonal lattice, harbor a positive (black) or negative (white) net magnetic charge, which can be revealed in the MFM (g) (Tanaka 2006).

### III.b Artificial honeycomb ice and its generalizations

Around the same time, Tanaka *et al*. (Tanaka 2006) performed a study on a continuous honeycomb network of ferromagnetic wires, which was published only a month after Wang's work (Wang 2006). Each vertex in this structure connects three nanowire links, and the magnetostatic energy of the vertex is minimized when it obeys a pseudo-ice rule which dictates that two moments point in and one points out, or vice versa.

This geometry and interaction is known as the honeycomb ice system, and was later closely studied both theoretically and experimentally (section V). Its various phases can be modeled as a hexagonal vertex model obeying the pseudo-ice rule, or as a kagome lattice of interacting dipoles, or as a hexagonal lattice of magnetic charges (Fig. 4). All these are useful descriptions of its various regimes(see below) and have therefore led to some degree of confusion in nomenclature, with the arrangement being called honeycomb, or hexagonal, or kagome spin ice. In this review we will stick with the intuitive name honeycomb that readily reflects its actual structure and declares no special description.

Tanaka and collaborators examined both the magnetic moment configuration and pioneered magnetoresistance measurements of these systems. Their measurements of magnetoresistance revealed sudden drops corresponding to spin switching, which suggested an ice-type interaction at the vertices of their honeycomb. To corroborate this intuition they performed an MFM analysis, which, however, in a network of connected nanowires, could only reveal the net magnetic charge at the vertices. These results were consistent with the hypothesis that the system obeyed the ice rules, although not without ambiguity, as demonstrated by Qi and collaborators (Qi 2008). They followed Tanaka's work with a study involving direct visualization via Lorenz microscopy (Qi 2008), which demonstrated rigid adherence to the honeycomb pseudo-ice rule after rotational demagnetization, as well as the absence of long-range order, in analogy to the square ice results. Qi and coworkers also found that the measured correlations were invariably of higher absolute value than the ones obtained from pure nearest neighbor modeling. The signs of the correlations were consistent with the effect of the long-range dipole interaction computed via magnetostatics.

In later years, a number of experimental studies explored other manifestations of artificial spin ice, both on connected networks and on separated islands. It is now possible to realize a variety of thermodynamic ensembles of artificial spin ice and to image them through a wide range of techniques. Of particular experimental interest have been the response of these systems to applied magnetic fields (Westphalen, 2008; Remhof, 2008; Schumann, 2010; Mengotti 2011; Huegli, 2012; Branford, 2012; Schumann 2012), issues of thermalization (Morgan 2011-2013), the nature of different lattice geometries (Ke, 2008b; Li, 2010a; Zhang, 2011), and the influence of disorder (Daunheimer, 2011; Kohli, 2011).

The contiguous arrays mentioned above (Tanaka 2006, Qi 2008) proved useful for further developments in magneto-transport measures (Branford 2012), although they pose issues in the direct visualization of magnetic spins through MFM imaging. On the other hand, arrays of separated nanoislands are readily imaged through MFM (Nisoli 2010, Li 2010a). Contiguous honeycomb ice is also associated with a lesser degree of disorder (Daunheimer 2011). By comparing experimentally measured spin correlations with the predictions of a model based purely on the nearest neighbor pseudo-ice rule, Qi also suggested evidence for the effects of long-range dipolar interactions in this system. Later we will discuss how theoretical descriptions of the honeycomb lattice that go beyond a purely vertex model predict lower entropy states, involving rearrangement of magnetic charges as well as an ordered loop state.

While square and honeycomb artificial spin ice differ both in topology and energetics, the brickwork lattice is intermediate between the two. It shares the topology of the honeycomb lattice while, as in square ice, the interactions between spins are not equivalent. From the point of view of pure vertex energy, it possesses a ground state, equivalent to that of the square artificial spin ice antiferromagnetic tiling. Li and collaborators compared the population of lowest energy vertices of brickwork artificial spin ice (after AC demagnetization) with that of both honeycomb and square artificial spin ice and found that brickwork behaves in ways much similar to the latter (Li, 2010a). That the local symmetry of the vertex might play a larger role than the topology of the array is perhaps not surprising: without a locally degenerate energy profile to start with, the global ground state is unlikely to exhibit extensive entropy.

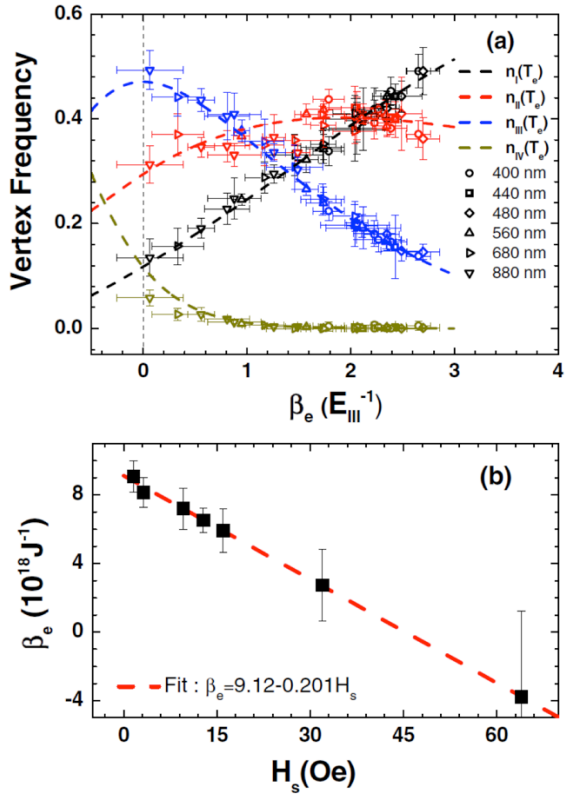

Figure 5. (Color Online.) AC demagnetization at different magnetic field step sizes returns widely different magnetic spin ensembles. The process can be seen as an external drive acting on a granular material, and its outcome can be described in terms of an effective temperature, which depends on the magnetic drive. Top: Relative frequency of different vertices of square ice (Figure 2) from square arrays of different lattice constants, annealed at different step size, are plotted against their effective reciprocal temperature. Lines show theoretical predictions based on an effective thermodynamics. Bottom: Linear dependence between the extracted reciprocal effective temperature and the magnetic field step size (Nisoli 2010).

A recent series of experiments (Zhang 2012) considered honeycomb and kagome lattices of magnetic moments pointing out of the substrate plane. As far as nearest-neighbor interactions are concerned, these evidently map onto an antiferromagnetic Ising model: the honeycomb is not frustrated while the kagome is. The out-of plane kagome, in particular, if described at the first nearest neighbors level, can be mapped into the in-plane honeycomb artificial spin ice. However, the profile of the long-range tail of the magnetic interactions is completely different: it is isotropic and leads to an effectively antiferromagnetic interaction at all distances $r$ and orientations (decaying as $r^{-3}$). Yet, quite strikingly, when Zhang and coworkers compared the pair spin correlation extracted from the two systems after AC demagnetization, they found that they almost matched. It seems therefore that field driven ensembles in these frustrated spin systems are dominated by lattice topology and nearest-neighbor interactions, with the long-range tail having little effect.

## IV. Characterizing and improving equilibration

The insight that the artificial frustrated arrays exhibit a non-thermal ensemble (Möller 2006) is at the root of two central conceptual questions that have since received much attention. Firstly, is there any way to equilibrate the arrays thermally? And, secondly, what is the novel physics involved in these intrinsically non-equilibrium systems? To these we turn next.

**IV.a Artificial spin ice as a non-thermal ensemble**

As mentioned above, in early experiments artificial spin ice was prepared via AC demagnetization: as frustration often leads to absence of magnetization, it appeared reasonable to demagnetize the arrays in order to reveal the underlying frustrated manifold. AC demagnetization has a long history, and it had been employed routinely to modify the domain structures in bulk ferromagnets. Artificial spin ice provided an opportunity to investigate how demagnetization relates to energy minimization at a constituent level. As the dominant interaction energy involves different islands sharing the same vertex, one can approximate the total energy of the array by assigning a vertex-energy to artificial spin ice ($E_{vert} = \sum n_i \in_i$ where $\in_i$ is the energy of a vertex of type i and $n_i$ its frequency, which can be extracted from MFM images) (Baxter 1982, Xe 2008). Thus, one needs to study the effect of AC demagnetization on the vertex frequencies, rather than just the macroscopic magnetization.

Early protocols of rotational demagnetization for artificial spin ice were able to disorder the vertices without in fact reducing their overall energy (Wang 2006, 2007; Nisoli 2007). Improved protocols could consistently return ensembles of desired energy, but it proved impossible to lower the energy of square ice beyond a limit far above the ground state (Ke, 2008). The most effective protocol found by Ke *et al.* involves rotating the sample in a magnetic field of alternating polarity and decreasing amplitude. The idea behind the protocol is that the external magnetic field anneals artificial spin ice only when it visits a certain window of opportunity, centered on the island coercive field, and whose size is given by the local field—controlled in turn by the lattice constant of the array. One expects that, by decreasing the amplitude of the steps, one can map that window ever more finely, hence (hopefully) perform a better annealing. A poorly-understood issue of the protocol is worth mentioning here: for optimal performance, the magnetic field needs to switch sign, which should be irrelevant given that the sample is rotating anyway. This suggests that peaks in the temporal rate of the magnetic field might be essential to the process and underlines the well-known fact that navigating complex energy landscapes is a highly non-trivial problem in itself.

Of course, direct inspection of microscopic degrees of freedom through MFM images provides much richer information about the microstate than its energy alone. Alongside the experimental effort, Nisoli *et al*. developed a simple theoretical treatment of AC demagnetization as a stochastic process (Nisoli, 2007, 2010), based on the assumption that AC demagnetized artificial spin ice can be treated as an externally driven granular material. Because of the complexity of its energy landscape and the fact that its microscopic degrees of freedom can be read directly via MFM scans, artificial spin ice can in principle provide a more robust and general validation for effective thermodynamics formalisms that have been more generally pioneered in studies of simpler granular materials (Behringer, 2002; Abate, 2008; Sollich, 1997; Colizza, 2002; Cugliandolo, 1997, 2011) as well as in the theory of topological defects in materials (Langer, 2010).

Assuming that the process of AC demagnetization can produce a well-defined statistical ensemble, the final microstate can be computed as the most likely outcome of that stochastic process. The optimization of suitable effective entropy, obtained by combinatorics and constrained to an energy manifold, defines an effective temperature and allows for prediction of the vertex frequencies (Fig. 5). The effective temperature can be controlled by the external drive and its reciprocal is linear in the amplitude of the magnetic step size with which the field is reduced (Nisoli 2007, 2010).

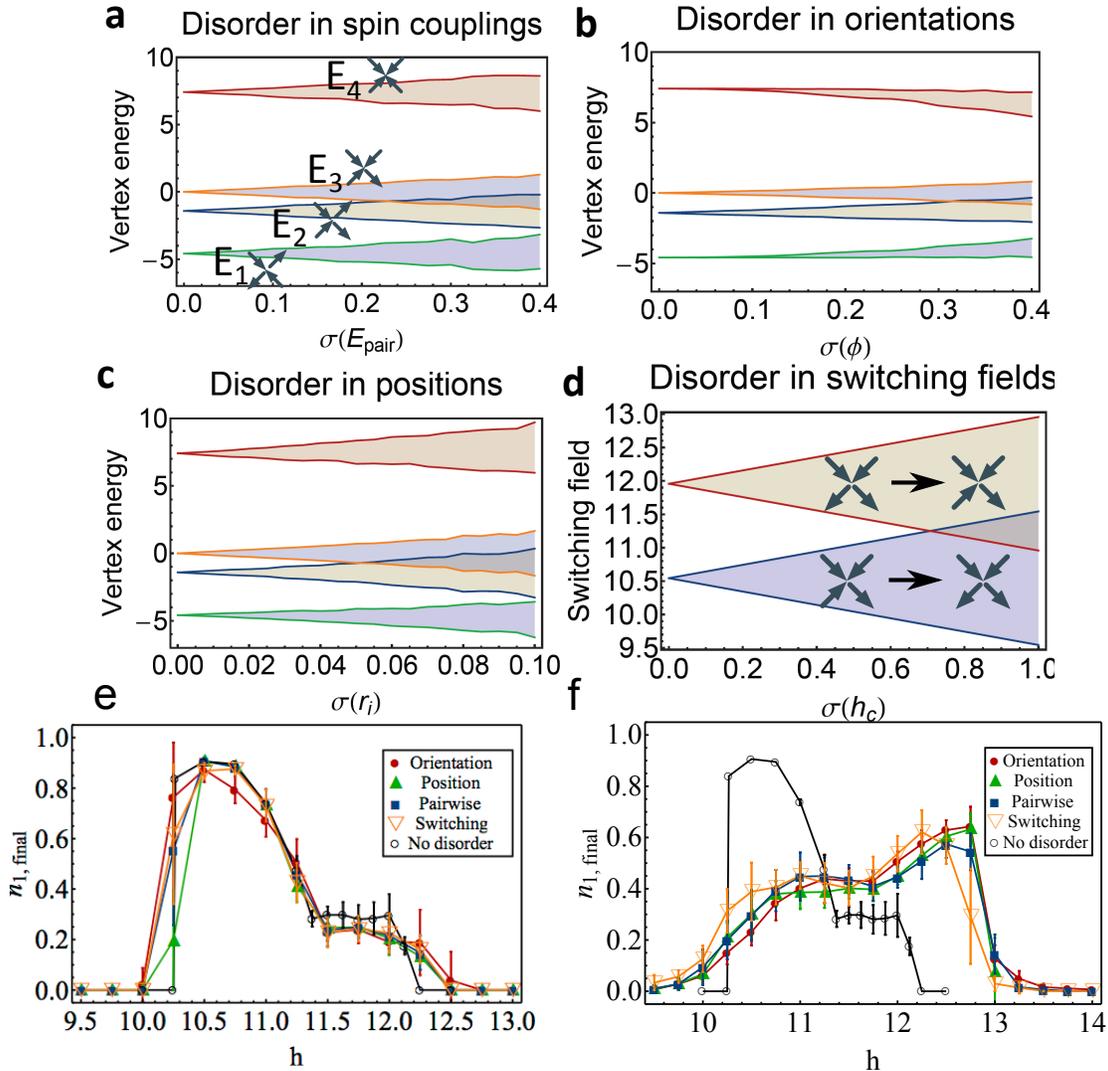

Figure 6. (Color Online.) Numerical simulations of the field-induced dynamics of artificial spin ice suggest that the strength of disorder is more relevant than its kind. The mean energy plus/minus one standard deviation of type IV, III, II, I vertices when disorder comes from to (a) pairwise energy, (b) orientation and (c) position. Crossing of Type II and Type III bands is taken as transition between weak and strong disorder. (d) The mean plus/minus one standard deviation of the external field required to convert a Type 2 to a Type 3 vertex (upper band) and the reverse (lower band). Plots of the relative frequencies of Type I vertices as a function of the applied rotational field (e, f) show that weak disorder accounts only for small perturbation (e), whereas strong disorder qualitatively changes the annealing profile, in a way independent from its kind (f) (Budrikis 2012).

## IV.b Kinetics and disorder

While controllable to a certain degree, any attempted magneto-agitation of square ice resulted in frozen disorder in which small grains of Type I vertices were visually discernible in MFM images. The portion of ground state vertices increased from a relative occurrence of 12% to a maximum of 55% (Fig. 5).

The kinetics of the process is still not fully understood, which is hardly surprising since the relevant phenomena include jamming, kinetic constraints, and glassiness. Clearly annealing out excited vertices cannot be done locally: two vertices of opposite magnetic charge need to meet so that they can annihilate in pairs. This amounts to a reaction annihilation problem supplemented by kinematic constraints. These constraints restrict the motion of monopole excitations towards one another, which can even get stuck to each other without being able to annihilate (Castelnovo 2010, Levis      2012).
Different computational studies have attempted to explain the failure to reach the square ice ground state via a variety of models of the demagnetization dynamics (Budrikis 2010, 2012a, 2012b, 2012c; Mol, 2009).

Disorder is a candidate cause for the lack of complete annealing of square ice through AC demagnetization. Islands have slight variations in shape, height, and relative position, which in turn translate into variations of the coercive field and  interaction energy. This effect is naturally more pronounced in artificial spin ice than in natural spin ice since it consists of lithographically-fabricated islands instead of identical atomic spins

Disorder in the coercive field of the nanoisland was revealed through magneto-optical Kerr effect measurements of the global magnetization of square arrays (Kohli 2011). Comparing the strong and non-monotonic angular dependence of the arrays' coercive field with micromagnetic simulations for arrays of different lattice constants, Kohli and collaborators concluded that global coercivity was strongly affected by a collective behavior rooted in the disorder of the islands' individual intrinsic coercive fields.  Islands with lower intrinsic coercive field activate cascades of reversals, which reduce the global coercivity  of the array. As expected, the effect was more pronounced at small lattice constant, where interisland interactions are stronger.

To explore the role of disorder in the coercive field and in the inter-island interaction, as well as of finite size effects on AC demagnetization, Budrikis *et al.* (Budrikis 2010, 2012a) carried out an extensive program of numerical investigation on square ice, although employing rotating fields *of constant strength*, in contrast to the stepped fields described earlier. They simulated the effects of a rotational magnetic field at different intensities close to the coercive field of the islands. The field switches spins randomly if the total magnetic field  (applied and local) exceeds the coercive field, and simulations run until a stationary state is achieved. Their findings seem to point to a major role played by the strength rather than the origin of the disorder (Fig. 6). In particular they identified a weak disorder regime in which the effect of disorder is only perturbative in the sense that the frequencies of the vertices are slightly altered while their dependence on the field is qualitatively unchanged, and a strong disorder regime in which the population of ground state vertices as a function of the applied rotational magnetic field is qualitatively altered from the perfect system.

While disorder is a good candidate to explain lack of annealing, recent more advanced equilibration techniques, discussed below, were shown to be capable of producing large crystallites of the square lattice ground state (Morgan 2011). We will see later that disorder plays a major role in nucleating spin flips and activating cascades in artificial spin ice under magnetization reversal (Mengotti 2011, Shen 2012).

### IV.c Artificial spin clusters

One way to understand the collective behavior of magnetic islands in artificial spin ice is to decompose it into its subsystems and study them as isolated nanoclusters. This can evidence peculiar properties of frustration, chirality or lack thereof in those constituents.

We have seen that AC demagnetization can be employed to lower the vertex energy of square ice, yet it never reaches its ground state this way. We also saw that, while it can access the kagome I phase, in which all vertices have charge +1 or -1, it cannot readily access any of the lower entropy phases induced by the long range of the dipolar interactions, (Möller 2009, Chern 2011), beside weak signatures in correlations (Lammert 2010, Qi 2008, Rougemaille 2010). To investigate the role that frustration plays in such mesoscopic systems, Mengotti *et al.* (2008) decomposed honeycomb artificial spin ice into clusters of one, two and three rings (Fig. 7), subjected them to AC demagnetization, and visualized the magnetic degrees of freedom via X-rays magnetic circular dichroism (XMCD) to study the configurations of lowest energy.

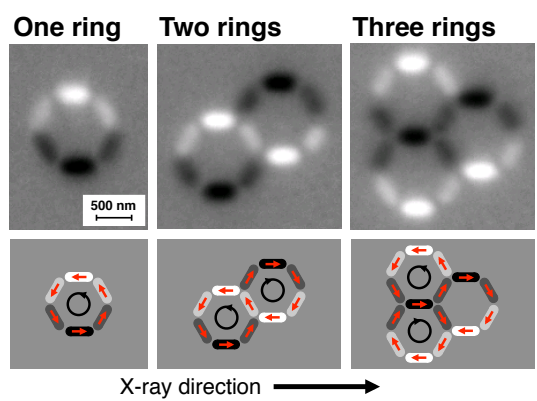

Figure 7. (Color Online.) Decomposing artificial spin ice into increasingly frustrated clusters can help understand the magnetic ensemble of an extended array. Here X-ray magnetic circular dichroism (XMCD) has been used to image the magnetic moments (Mengotti 2008).

Clearly, the one-ring clusters are non-frustrated. Perhaps not surprisingly it was found that at sufficient dipolar strength almost all (94%) of the cluster moments were arranged in the head-to-tail loop configuration that comprises the lowest energy configuration. Once one adds more clusters, three-legged frustrated vertices appear. Experiment shows that the ground state was reached only in about half (48%) of the two ring clusters and one third (31%) of the three ring clusters, corresponding to a suppression of the lowest energy configuration. As for extended arrays, AC demagnetization can always anneal the nearest neighbor interaction at the vertex level, so the pseudo ice rule is always respected. But as the clusters grow in size, the ability to reach the lowest energy configuration dictated by the comparatively weak energy differences due to the long-range part of the dipolar interaction is lost. These results are consistent with findings from AC demagnetization of the extended honeycomb lattice, in which the pseudo-ice rule manifold is always reached yet the non-degenerate ground state, in which the degeneracy is lifted by long range dipolar interactions, is never achieved. These insights are also relevant for recent experiments on magnetotransport (Branford 2012, see below), in which chirality is suspected to arise from the formation of loop configurations at the edge of the array.

Li *et al.* (Li 2010b) also investigated the lack of annealing into the ground state in square ice clusters and its possible relationship with frustration and jamming in AC demagnetization. They suggested that a kinetic bottleneck between near-degenerate states—rather than only strong disorder—is responsible for the lack of annealing in the extended square arrays.

### IV.d Second generation equilibration schemes

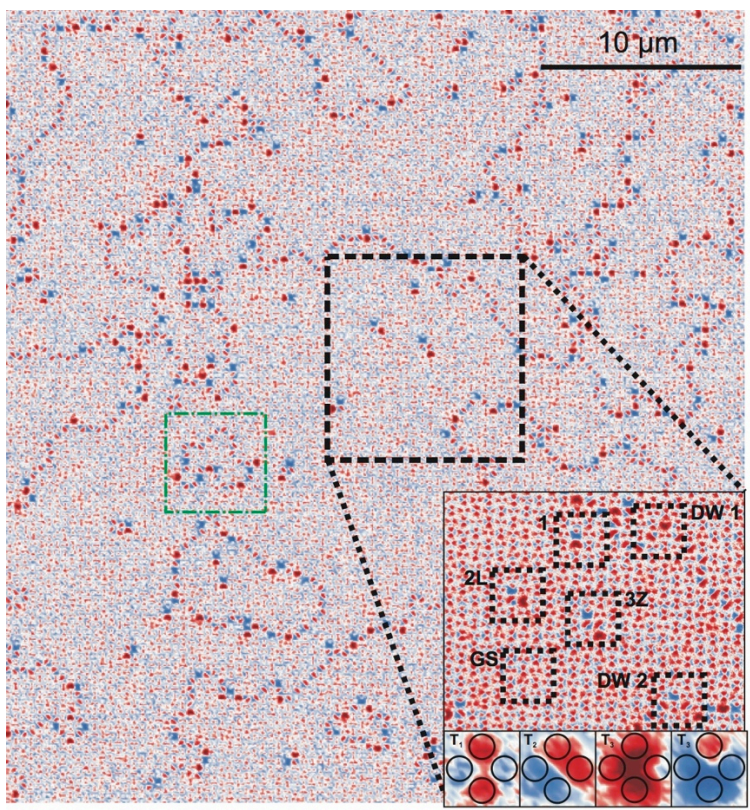

Figure 8. (Color Online.) Large domains of the ground state of square artificial ice can be reached via during growth, as it can be seen in this MFM image: North and south poles are represented by red and blues. Antiferromagnet-like domain boundaries are visible as well as localized elementary excitations. The magnified region shows elementary excitations ion the ground state (type I, II and III, see also Fig. 2) (Morgan 2011).

Since large interisland interactions ($10^4$-$10^5$ K) and intrinsic coercive fields ($10^2$ Oe) make artificial spin ice insensitive to thermal fluctuations, most of the early experiments involved subjecting the arrays to an oscillating magnetic field. As explained above, this method, widely employed for material demagnetization and therefore disordering, was optimized to produce a certain degree of order at the vertex level. However, a finer control over the magnetic ensembles, including the ability to equilibrate them in different thermalized states, is still highly desirable. After all, these systems were designed initially to replicate celebrated models of statistical mechanics.

Clearly, high coercive energies are a consequence of the nanoscopic size of the islands. A route to attack this problem is to notice that the islands thermalize as they grow during fabrication. One assumes that when the islands are below a critical height (which is dictated by temperature), artificial spin ice visits an energy region in which the magnetic interactions and the coercive energies are of the order of the thermal energy or smaller; then their spins should be thermally equilibrated. This was in fact realized experimentally by Morgan *et al.* (2011). They grew a square lattice of permalloy nanoislands on a silicon substrate. When island growth was complete, MFM imaging revealed that the island magnetic moments were effectively frozen into a patchwork of large crystallites (of ~10 micrometers or about 25 lattice constants) of chessboard tilings of Type I vertices (the square ice ground state) separated by grain boundaries of Type II and III vertices (Fig. 8). Growth-induced equilibrium seems validated by the observation of a Boltzmann distribution in the sparse vertex excitations inside the domains. Interestingly, the authors observed a scarcity of excitations consisting of a pair of Type III vertices separated by a long, straight string of Type II vertices and an excess of closed loops of Type II vertices—a phenomenon they attributed to attraction between defect pairs.

Since the magnetic degrees of freedom of an as-grown thermalized array seem to form an equilibrated ensemble, it should be possible to control the effective temperature by tuning thermodynamic parameters at fabrication. It was suggested theoretically (Nisoli 2012) and then confirmed numerically (Greaves 2012) that ensembles of lower entropy might be achieved—somewhat counter-intuitively—by raising the temperature during deposition, as fabrication at higher temperatures would extend the dynamical range during growth. In particular, if the geometry is conducive to a phase transition, then there will be a temperature threshold at fabrication above which the lowest energy phase is reached. Morgan's technique could then allow the exploration of the numerically predicted but as-yet unobserved magnetic-charge-ordered and ground state phases of the honeycomb lattice (Möller, 2009; Chern 2011) (Section V.c.).

The as-grown technique returns a non-dynamical, frozen-in ensemble of spins and in this sense, it is via spatial self-averaging of the system that the statistics of a Gibbs ensemble arises. This can be probed by MFM imaging. In the presence of (small) external fields, phenomena such as dissociation of magnetic charges or string avalanches starting from a thermalized ensemble can be investigated — something that has not been attempted so far.

On the other hand, as the experimental search for magnetic monopoles in natural spin ice proceeded from monopole observation (Jaubert, 2009; Morris, 2009; Bramwell, 2009; Fennell, 2009) to measurements of "magnetricity" (Giblin 2011), a newer, more dynamical artificial spin ice is required to implement effects associated with monopole propagation, and as a potential medium for magnetic circuitry.

Dynamical artificial spin ice can be realized by δ-doping Pd (Fe) to engineer nanoislands of low Curie temperature, as recently demonstrated by Kapaklis and coworkers (Kaplakis 2012). They performed measurements of magnetic hysteresis loops using the magneto-optical Kerr effect (MOKE) at temperatures ranging from 5 to 300 K, both on a continuous film and on a dynamical artificial spin ice of square geometry. Because of the applied field, the energetics of the vertices shown in Figure 2 changes: Type II vertices, polarized along the field, become energetically favorable, leaving as a ground state a magnetized tessellation of Type II vertices, with Type I vertices as excitations. This leads to a remnant magnetization in the hysteresis curves, which is indeed what is found at lower temperature.

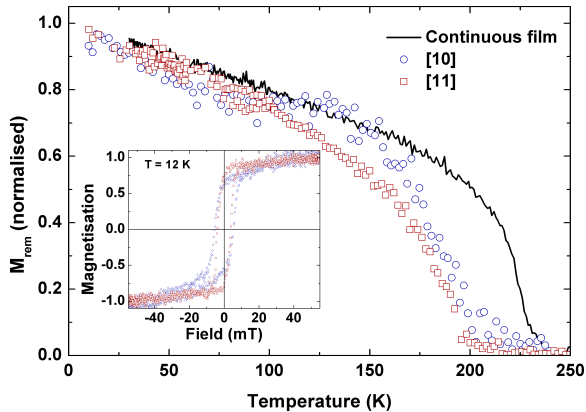

Figure 9. (Color Online.) MOKE hysteresis cycles performed at different temperatures on square artificial spin ice made from a material of low Curie temperature, show the disappearance of the remnant magnetization at temperatures below the Curie point. This signals a "melting" of the macro-spin degrees of freedom in square artificial ice, likely due to thermally activated flips of island macro spins, in thermally activated dynamics. The inset shows representative normalized magnetic hysteresis loops for a magnetic field applied parallel to the [10] and [11] directions at a temperature of 12 K (Kapaklis 2012).

Yet as temperature grows to about 200K, the magnetization of the δ-doped Pd(Fe) square ice is strongly reduced and rapidly drops to zero *before* the Curie temperature (230K) of the thin film of the same height (Fig. 9). The authors interpret this result as melting of the ground state due to thermal excitation of energetically unfavorable vertices whose magnetization is either zero (Type I), or at least not aligned to the applied field. Of course one must exclude melting of the individual island moment to invoke collective melting of artificial spin ice. In a later work the authors added direct visualization to their technique to exclude that possibility (Arnalds 2012).

These new methods present the possibility of producing thermal ensembles for the artificial spin ices. Such thermal ensembles are essential for the study of magnetic phases of honeycomb ice and the lower temperature monopoles and more. The two approaches can be mixed and matched: as mentioned, square artificial spin ice could be prepared *à la* Morgan, in a frozen thermalized ensemble, and then probed with fields, to study, e.g., monopole pair dissociation. Alternatively, in a dynamical artificial spin ice, *à la* Kapaklis, the temperature could be fine-tuned to investigate the response to external fields in a glassy regime of slow dynamics.

## V. True degeneracy, monopoles and more

### V.a The quest for true degeneracy

The asymmetry of the dipolar interaction between collinear and perpendicular islands converging into the same vertex (Fig. 2) endows artificial square ice with a well-defined and non-degenerate ground state. This same asymmetry is clearly absent in the four-legged vertices of the three dimensional pyrochlore lattice of spin ice (Fig 1). The asymmetry of the two dimensional case translates into a preference for Type I vertices and the resulting ground-state order. How can the frustration-induced degeneracy be preserved down to lower temperatures or stronger couplings?

A simple strategy to make square ice degenerate was proposed early on (Möller 2006). A symmetric arrangement of four points requires a third dimension (i.e., a tetrahedron), Figure 10. Therefore, it was proposed to generate a quasi-three dimensional array by placing islands pointing in the x- and y-directions on different heights, so that their endpoints do form a tetrahedron. As the endpoints of the islands approach each other, the ice rule degeneracy becomes increasingly accurate, despite the presence of complex further neighbor interactions. This is also interesting as it would be a first step towards a three-dimensional version of artificial spin ice, which has not been attempted yet. (This model was further studied numerically by Mól and coworkers (Mól 2010).)

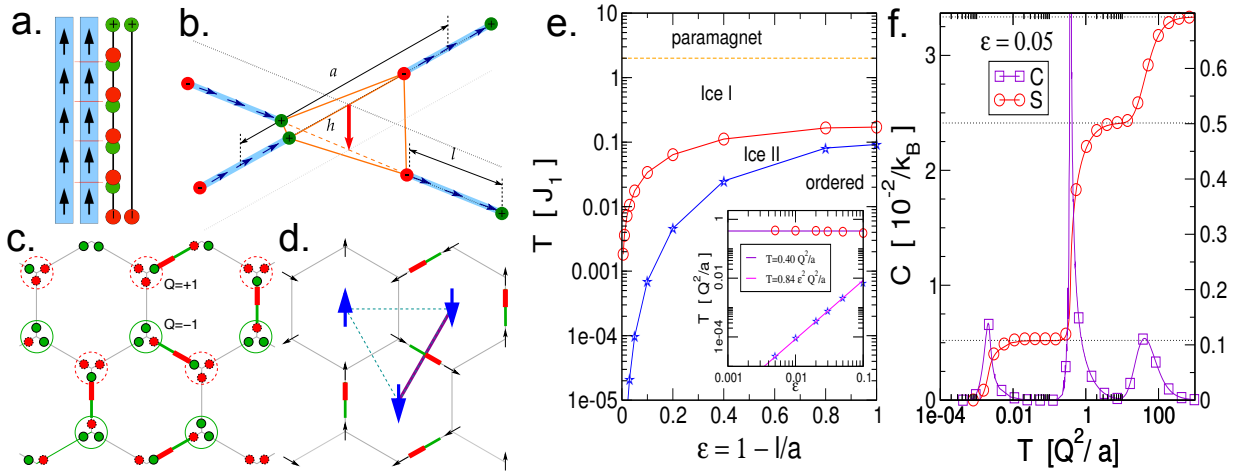

Figure 10. (Color Online.) a: A thin, uniform dipolar needle of moment $\boldsymbol{\mu}$ and length $\boldsymbol{d}$ is equivalent to a pair of magnetic charges of equal and opposite strength $\boldsymbol{\mu/d}$ at its ends: pictorially, the needle is a string of tiny bar magnets whose adjacent charges cancel everywhere except at the ends. b: In a square geometry when the endpoints of the needles sharing a vertex are much closer to each other than the needles' length, the energy of a configuration is dominated by the interaction between nearby charges. Introducing an appropriate height offset $\boldsymbol{h}$ between perpendicular needles places these charges at the corners of a tetrahedron, whose high symmetry renders Type I and II ice vertices approximately degenerate. c: For honeycomb ice, an odd number of needles share a vertex, and vertices necessarily carry nonzero, odd charge. d: As vertices are not rotationally symmetric, they also carry a dipole moment with respect to their midpoint, denoted by the black arrows. In Ice I, charge $|q|=3$ vertices are energetically penalized and vertices have (predominantly) charge $|q|=1$. The long-range dipolar interaction generates Coulomb interactions between the charges on different vertices, leading to simple charge ordering characteristic of Ice II: charges +1 reside on one sublattice of the honeycomb lattice, and charges -1 on the other. Ice II still has nonzero entropy, related to that of a triangular lattice dimer model (d), as described in the text. Interactions between the vertex dipole moments drive a final ordering transition that breaks translational symmetry. e: the phase diagram as a function of $\varepsilon = 1 - l/a$, where $l$, is the island length, and $a$ is the lattice constant, and with temperature in units of the nearest neighbor interaction constant $J_1$. f: In the limit of small $\varepsilon$, the four different thermodynamic regimes (paramagnet, Ice I, Ice II and ordered) are identifiable through plateaus in the entropy (Möller 2009).

A different approach towards generating stable degeneracy does not involve the third dimension, which might present problems in nanofabrication, but instead invokes a naturally degenerate degree of freedom. Rather than engineering the degeneracy in the vertex energetics, one can use non-degenerate vertices, yet arrange them in such a way that frustration and therefore degeneracy results from the impossibility to place all of them simultaneously in their lowest energy configuration (Morrison 2013). Besides residual entropy, such "vertex-frustrated lattices" seem—at least theoretically—to be able to access novel exotic states, such as smectic phases and sliding phases.

To date, the only experimentally attempted version of a truly degenerate artificial spin ice is the honeycomb lattice described above. Here degeneracy is achieved at the vertex level when the islands are arranged along the bonds of a honeycomb lattice, so that their midpoints form a kagome lattice (Fig. 4). The ground state of a vertex is degenerate, since all three "legs" converging in a vertex at the same relative 120° angle and thus have equivalent interactions. The ground state is characterized by the pseudo-ice rule, a 2-in/1-out or 2-out/1-in rule that provides maximum satisfaction of the frustrated interactions (Möller 2006). For the full lattice, this is strongly reminiscent of the actual ice-rule on a pyrochlore lattice, but the change from tetrahedra to triangles is less innocuous – and leads to richer physics – than it at first appears.

As a net magnetic moment and indeed even a net magnetic charge can be assigned to the odd-legged vertices, the long-range interactions (Fig. 10) between these give rise to four regimes which were recently investigated theoretically and numerically (Möller, 2009; Chern 2011) as described in detail above (Fig. 10): a high-temperature paramagnet, the Ice I and Ice II phases already mentioned above, as well as a low-temperature ordered state. The second-generation equilibration schemes might be employed to investigate the kagome Ice II and loop phases, which have not yet been experimentally observed.

### V.b Monopoles and multipoles

Despite much searching (Goldhaber 1990, Particle Data Group 2012), no convincing experimental evidence has emerged for the existence of elementary monopoles. To explore how monopole excitations appear in artificial spin ice, we first consider what is meant by an emergent magnetic monopole.

Consider first the very simple idea of a magnetic charge: something that sets up an appropriate divergence in the magnetic field, leading to a magnetostatic interaction in the form of a Coulomb law. One can idealize the island-shaped real dipole as a thin uniform needle (Fig. 10 a.), which is in turn equivalent to a dumbbell of moment $\mu$ and length d and therefore of equal and opposite magnetic charge of size $\mu$/d (Jackson 1998). These are monopoles only for bookkeeping purposes, *i.e.*, to compute energies and fields, not a priori related to any collective low-energy physics. As the endpoints of the islands are close to each other, the leading term in the energy would like the endpoints of the islands impinging on a joint vertex to have a minimal total charge. This motivates a change of variables: instead of considering dipolar islands on the links of a lattice, one can analyze the energetics conveniently as charge distributions on the *vertices* of the lattice. While modeling not dipolar islands, but atomic dipoles in the rare earth spin ice compounds, Castelnovo et al. (Castelnovo 2008) proposed this dumbbell model. It was already known that defects violating the ice rule naturally live on the vertices of the lattice (Moessner 2003), and Ryzhkin pointed out that such defects coupled to an applied magnetic field as if they carried a magnetic charge (Ryzhkin 2005).

The dumbbell model demonstrated that these defects interact via a magnetic Coulomb law (Castelnovo 2008) and can be separated with finite energy expenditure: they are genuine deconfined fractionalized excitations--the emergent magnetic monopoles.

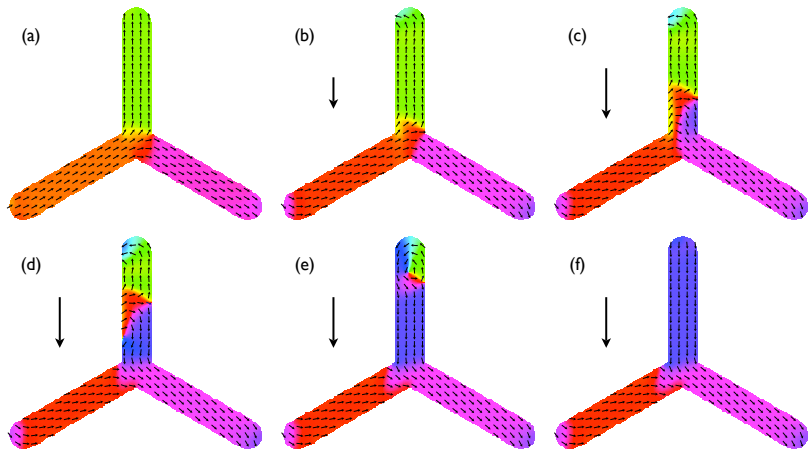

Figure 11. (Color Online.) Inside a vertex element of artificial spin ice realized by connected wires. Magnetization first adjusts adiabatically (panels a-c) to an applied magnetic field. When the applied field reaches a critical value, the reversal proceeds via propagation of a detached domain wall (panels c-f) while the field value remains essentially unchanged (Shen 2012).

Thinking of each bond carrying a unit of flux, the ice-rule implies that the flux field so defined is divergence-free.

This emergent conservation law lies at the base of what makes these ice models interesting. Incidentally, it also establishes the connection to gauge theories, whose link variables – electric fluxes in that language – in turn satisfy Gauss' law, div $E = \rho$, and magnetic monopoles are then located at vertices violating the ice rule.

The reason they are *magnetic* monopoles is related to the fact that the link variables carry real magnetic moments in natural and artificial spin ice. If they were built of electric dipoles, the defects would instead be electrically charged. Magnetic monopoles capture precisely this leading-order physics in natural spin ice materials. Yet sub-leading terms also play a crucial role in lifting degeneracies, as we will see below – formally, these correspond to higher-order multipoles.

In the case of even lattice coordination (*e.g.*, the square lattice), imposing the ice rules amounts to demanding charge neutrality on each vertex (as verified in numerical work); while for odd coordination, each vertex harbors an odd charge, i.e., at least $\pm q$. Here, the magnetic charges appear as degrees of freedom, capable of simplifying considerably the low temperature energetic description and providing an *efficient* way of bookkeeping.

There now are the further demands on *bona fide* monopoles as quasiparticles: they must exist as independently mobile dynamical degrees of freedom, e.g. as sparse excitations which can move on top of a background of vertices satisfying the ice rules; and they must interact through a Coulomb potential as in the case of the rare earth spin ice compounds mentioned above—for a review, see Castelnovo 2012.

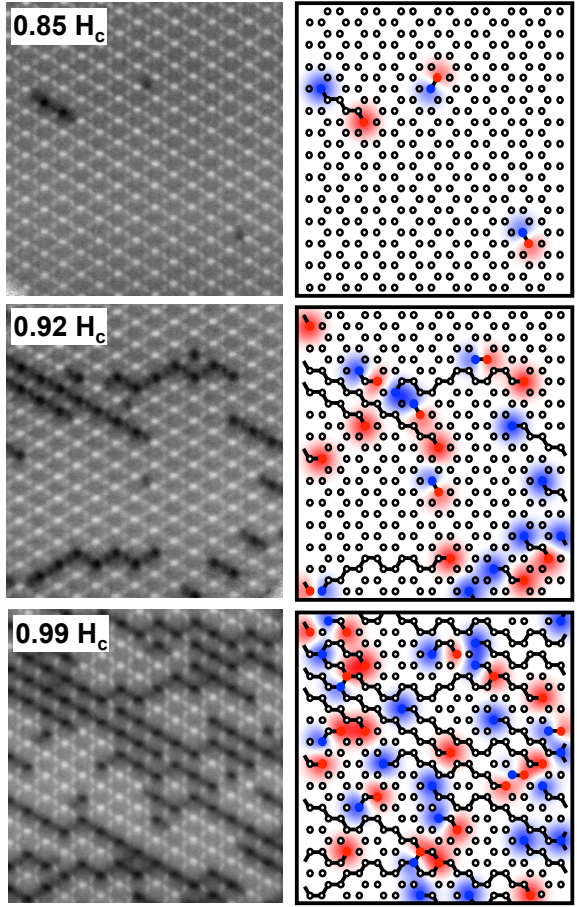

Figure 12. (Color Online.) Field reversal in a fully polarized honeycomb lattice proceeds by creation and propagation of avalanches of pairs of opposite magnetic charges separated by a Dirac string, observed directly via XMCD (left panels). On right panels we have the associated $\Delta Q$ map, showing the progressive change of magnetic charge on vertices. From top to bottom are shown progressive reversal at applied field which is 85% to 99% of the coercive field $H_c$ (Mengotti 2011).

The situation is considerably more complex in the case of artificial spin ice. (For completeness, we note that in thermal equilibrium, monopoles experience also an emergent entropic, *two-dimensional, logarithmic* Coulomb interaction, which therefore in principle precludes deconfinement anyway. (The interplay between magnetic and emergent entropic charge is discussed in Moessner 2010.)

For square ice, the ice rule corresponds to charge neutrality, yet the asymmetry of the magnetic interaction in 2D lifts the degeneracy, and as discussed above provides a unique antiferromagnetic ground state of Type I vertices (Figure 2). Flipping a spin in the ground state produces two nearby opposite magnetic charges (Type III vertices), but further flips to separate them generate a string of excited vertices (Type II). This string is endowed with a thermodynamic line tension, essentially arising from the energetic cost of the defects it consists of, but renormalized by thermal fluctuations, as verified in numerical work (e.g. Mól 2009). In the absence of an experimentally thermal ensemble in the artificial spin ices, a disordering transition in which the string loses its tension has not yet been witnessed experimentally.

The case of honeycomb spin ice differs yet further (Fig. 10). Because of the odd coordination number, a single vertex cannot carry zero magnetic charge, and instead has to have q = ±1 or ±3. It should be noted that the artificial spin ice literature follows a convention whereby such entities – vertices with nonzero magnetic charge – are also frequently referred to as magnetic monopoles. In the materials physics/exotic magnetism literature, by contrast, the stricter definition mentioned above of emergent fractionalized *deconfined* quasiparticles interacting via a magnetic Coulomb interaction is more frequently used to make a distinction with the 'bookkeeping' magnetic charges. (On top of all this, the term (magnetic) monopole is also used abstractly in the gauge theory literature without any direct reference to Maxwell electromagnetism.)

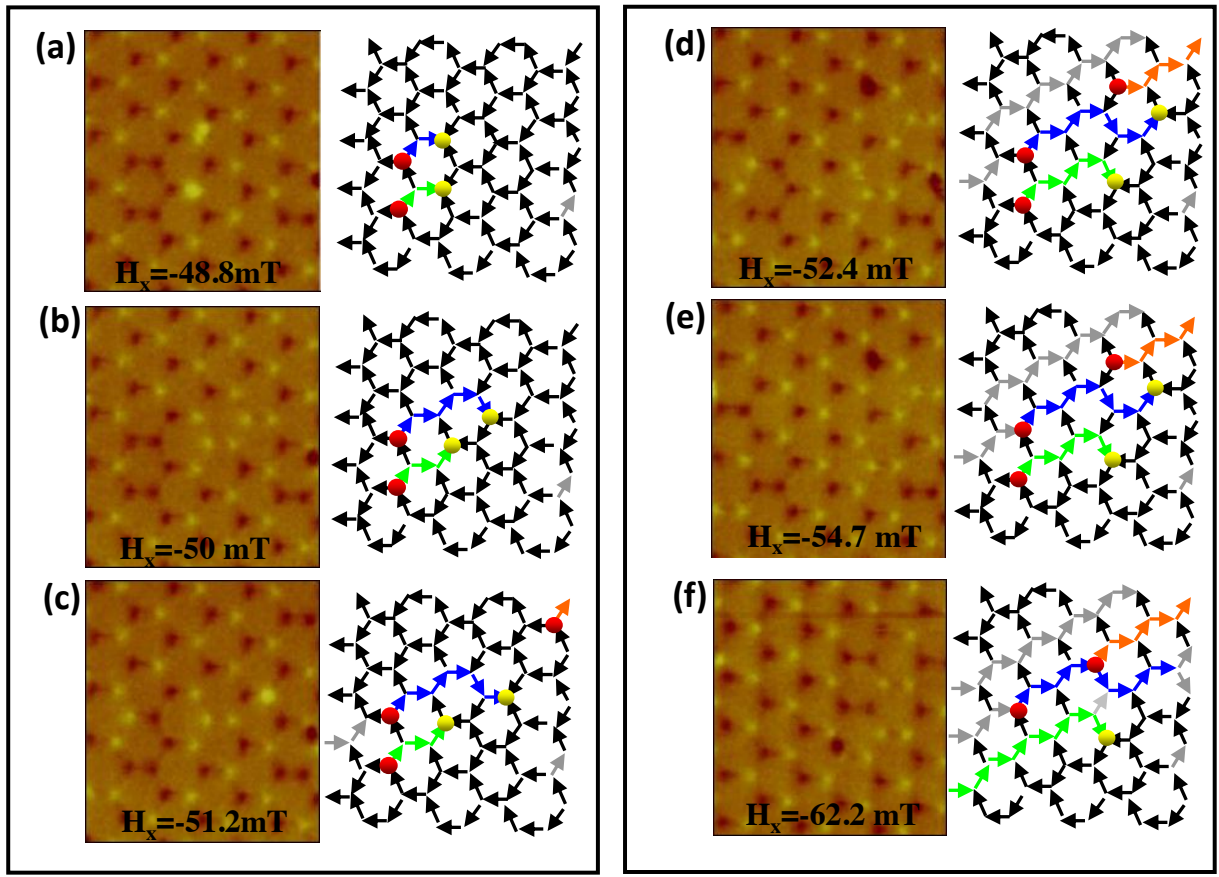

Figure 13. (Color Online.) Experimental observation of magnetically charged defects and their motion through the system in the process of magnetic reversal in cobalt honeycomb artificial spin ice: MFM images rendered as arrow cartoons (left and right panels, respectively) as the field strength is increased. (a) Two defect pairs with |ΔQ|=2 shown in yellow (Q=3) and red (Q=-3). Colored arrows denote the strings of flipped spins. (b) Motion of the magnetic charges involves extending the strings, with the field pushing red (yellow) defects to the left (right). (c) Arrival of a new red defect, (d) which moves left (f) until it is blocked by the string left behind by another defect (Ladak 2010).

As emphasized above, all $|q| = 1$ (in units of $\mu$/d) vertices are degenerate in isolation, but this degeneracy is lifted by long-range interactions as described below. The pseudo-ice rule manifold (two in, one out and vice versa) arises upon cooling when the energetically costly all-in or all-out vertices disappear. This Ice I regime is effectively a collection of magnetic charges q = ±1, which can achieve its lowest energy when opposite charges crystallize on neighboring vertices in a manner similar to the cations and anions in rock salt.

The charge-ordered regime, Ice II, retains a non-zero residual entropy since many microstates are consistent with a given charge state. Their number can be enumerated by a mapping to the exactly soluble dimer model on the honeycomb lattice (Möller 2009): by coloring in each island which contributes the minority charge -q on a vertex of net charge q at its ends, one obtains a hardcore dimer covering (Fig. 10). The degeneracy of Ice II is further lifted because the vertices not only have a net charge

|q|=1 but also higher-order multipoles. On account of the lack of rotational symmetry of the charge distribution on a given vertex, the dipole moment $v$ with respect to its center points along an axis joining the vertex to its minority charge (black arrows in Fig. 10 d). Note that the dipole moment $v$ of the vertices is distinct from the dipole moment $\mu$ of the islands.

The dumbbell model can thus be extended to include not only the leading Coulomb interaction between the charges, but also the full multipole expansion of the interaction energies of different islands (Möller 2009). In this multipole expansion, one encounters the following energy scales. Firstly, $E_\sigma$, the on-site vertex energy; next, $E_c = \mu_0 q^2/4\pi a$, ($a$ is the lattice constant) the strength of the nearest-neighbor interactions between magnetic charges; and finally, $E_d = \mu_0 v^2/a^3$, the corresponding dipolar energy scale. In the case where the magnetic islands almost touch at the vertices, i.e. when their length $d$ equals $(1-\varepsilon)a$ , one can construct a controlled perturbation theory in $\varepsilon$, with $E_\sigma \sim O(1/\varepsilon)$, $E_c \sim O(\varepsilon^0)$ and $E_d \sim O(\varepsilon^2)$ , since the dipoles have strength $|v| \sim \varepsilon a q$.

The phase diagram is shown in Figure 10. For T >> $E_\sigma$, the system is a conventional high-T paramagnet. For $E_\sigma$ >> T >> $E_c$, vertices of charge ±3 are expelled, reaching (Wills') kagome ice, Ice I. This phase is, from a symmetry perspective, identical to the paramagnet. For $E_c$ >> T >> $E_d$, magnetic charge ordering occurs, via an Ising transition, resulting in a state of charges +1 on one sublattice of the honeycomb lattice formed by the vertices, and charges -1 on the other. This is the second kagome ice, Ice II. Finally, as $E_d$ >> T, the dipoles order by a transition breaking translational symmetry, relieving the system of its remaining entropy. Note that, a similar ordering transition is observed in Monte Carlo simulations of (but not in experiments on) the spin ice compounds (Siddharthan 1999 and 2005, Melko 2001). This is due to corrections (Isokov 2005) to the dumbbell model, which are either due to multipolar or superexchange effects. Similarly, the ordering in artificial square ice can be thought of as a deviation from the ideal degenerate case (Möller 2006, Möller 2009).

Away from the limit of small $\varepsilon$, these scales are no longer well separated, and the resulting phase diagram is rich (Chern 2011). Whereas the charge ordering always occurs via an Ising transition, the phase transition at lower temperature turns from Kosterlitz-Thouless into three-state Potts as defects in the charge order destroy the algebraic correlations of the intermediate phase.

In the presence of a sufficiently strong applied field, the ground state turns out to be a different one: like in an ordinary magnet, the dipoles align, and one ends up with the maximally polarized state (Möller 2009, Chern 2011). Such ground state selection is well known in the rare earth spin ices (Moessner 2003) as even a uniform field couples non-trivially to the moments on account of their non-collinear axes (Moessner 1998). This turns out to be highly relevant to the monopole experiments described below.

As mentioned above, the kagome ice II and ground state phases have not yet been observed experimentally in honeycomb spin ice, although subtle suggestions of magnetic charge correlations have been reported in athermal ensembles obtained via AC demagnetization (Qi 2008, Lammert 2010, Rougemaille 2011). Clearly better annealing techniques are needed to explore the phase space for honeycomb ice. In general, two complementary experimental strategies can be attempted. One is to consider thermalized model systems, and work toward their experimental realization by building on methods described in Section IV.b; the other is to look for signatures of mobile magnetic charges in the athermal ensemble, "non-equilibrium monopoles," as in the experiments described next.

### V.c Collective physics in honeycomb artificial spin ice

We can now move on to the experimental monopole search, taking a short detour into the basics of the dynamics of the arrays. It has certainly been a pleasant surprise that the frustrated microarrays provide a new instance of non-equilibrium physics, one that can be manipulated externally while being probed microscopically. Our aim will therefore be to capture the observed non-equilibrium dynamics in at least a semi-quantitative way, in two steps: understanding the elementary dynamical process of a single moment reversing its magnetization (Fig. 11); and the collective dynamics of such moments reversals, which turn out to give rise to avalanches (Fig. 12, 13) with rather special properties, on account of the appearance of one-dimensional 'Dirac strings' as natural degrees of freedom. It is at this point the community of artificial spin ice is at its most interdisciplinary, combining micromagnetics with the study of disordered systems, real-time non-equilibrium dynamics and more.

Shen and coworkers have discussed the flipping of a single moment in continuous honeycomb networks in considerable detail. We do not expose here the dynamical process itself, which affects the flipping, but do note that, in the effective picture of dipolar needles underpinning the dumbbell model, it has an attractively simple description (Shen 2012). The flipping process consists of the positive (say) end of the needle emitting a charge +2q leaving behind a negative charge –q to satisfy charge conservation (Fig. 11). The emitted charge travels down the needle, and finally combines with the negative charge at the other end, leaving behind a flipped moment, and potentially inducing the next moment to flip. This process requires an activation energy, which depends on the detailed properties (such as geometry and anisotropy) of the structure under consideration.

With this in hand, let us consider the pair of prominent experiments (Ladak, 2010; Mengotti 2010) which studied magnetization reversal in order to unearth real-space evidence of the non-equilibrium monopoles, as well as to provide detailed insights into the non-equilibrium process of magnetization reversal. The basic idea is to prepare the system in a saturated state by applying a field and then monitoring the response to a reversed field. This problem is well studied in the case of a conventional ferromagnetic material, where a one-dimensional domain wall that separates the oppositely-magnetized domains is swept across the system. In contrast, for systems in the class considered here, it is strings, one-dimensional objects (Kasteleyn 1963), which effect magnetization reversal, since the maximally polarized ice state is not connected to other ice states by local island flips (Moessner 2003). Indeed, as a field drives non-equilibrium monopoles of opposite charge in opposite directions, complex energetic and topological interactions emerge between them, the strings they create, and the polarized background.

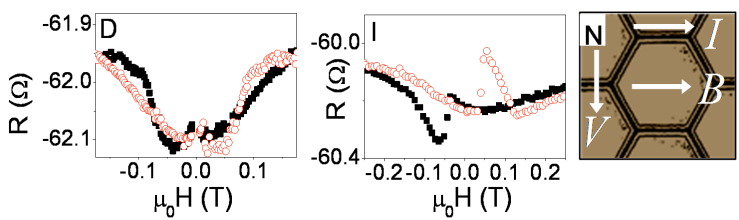

Figure 14. (Color Online.) Raw magnetotransport data on a cobalt honeycomb spin ice from (Branford 2012). Resistance versus field with positive (filled squares) and negative (open circles) sweep direction. Data were taken at 100 K (left) and 2 K (middle); on the right, schematics of the measurement geometry, which eliminates the conventional Hall effect. Note the unusual peak emerging at low temperatures. No such component is observed in an unpatterned film. (Branford 2012).

The achievement of these experiments is to image not only those strings but also their endpoints, at which the charge pattern breaks that established in the transition to the ice II phase: magnetization reversal changes both mono- and dipolar ordering. In analogy to spin ice (Castelnovo 2008), these charged excitations were termed monopoles, and the string of reversed dipoles Dirac strings. Ladak and coworkers (Ladak 2010) were the first to present a study of their field-driven motion using an MFM to image a nanostructure of cobalt wires arranged in honeycomb patterns. Such arrangement reveals the $\Delta Q = \pm 2$ charges at the end of each string on top of the $\pm q$ charged background. Performing imaging via synchrotron X-ray magnetic circular dichroism (XMCD) on a permalloy array of nanoislands, Mengotti and coworkers (Mengotti 2010) provided direct visualization of both the charges and their Dirac strings (Fig. 13) and revealed processes of nucleation and dissociation through string-avalanches.

Crucially, the detailed information on the profile of magnetization reversal provided a first semi-quantitative handle on the disorder present in the artificial spin ices. By monitoring magnetization profile and defect density, it was possible to compare experimental results to microscopic numerical simulations that found satisfactory agreement with simple models of the disorder with typically only a fitting parameter for the overall disorder strength (Ladak 2010, Mengotti 2011).

Later, Daunheimer and coworkers offered a more sensitive method to directly extract the distribution of coercivities directly from experimental data, without relying on Monte Carlo simulations. They applied the reversal magnetic fields at 120° and 100° in the direction of the initial polarization, rather than 180°, conditions in which magnetization reversal proceeds independently of their neighbors and without formation of avalanches. An interesting finding is that a connected honeycomb ASI, such as the one employed by Tanaka (Tanaka 2006), shows much smaller coercive disorder (Daunheimer 2011).

While the application of an external field is a clever way to overcome the coercive barrier toward spin flipping, an *in vivo* feedback of the structural changes is clearly desirable. Branford and coworkers recently pioneered one approach: in coincidence with field sweep, they performed measurements of magnetotransport, in particular monitoring the Hall signal (Branford 2012). In measurements of magnetoresistance in the harmchair direction of the honeycomb lattice, with magnetic field B pointing in the same direction and voltage V along the zigzag direction, they observed an anomalous Hall signal at temperature below 50 K, in the field range associated with flipping of the magnetic moment (Figure 14). Such signal is absent in non-patterned films. Their results point to the spontaneous appearance of chirality as the islands flip. In the absence of out-of-plane field or magnetization or a relative nonzero angle between the field and average current direction, this signal should vanish by symmetry. Their micromagnetic simulations point toward the formation of oriented loops of island moments at the edge of the sample as the origin of the nonzero chirality. This seems consistent with the numerical results for the Ice II phase of honeycomb ice as it approaches the ordered state (Chern 2011, Möller 2009), as well as with Mengotti's experiments on honeycomb nanoclusters (Mengotti 2008).

Taken together, the works described in this section have provided a pleasingly complete picture of dynamics and thermodynamics of honeycomb artificial spin ice, including its unconventional emergent degrees of freedom, while indicating a generic route to exotic states in artificial spin ice.

## VI. Other "artificial" spin systems

In addition to the artificial frustrated magnet systems composed of magnetic materials, there have been substantial efforts in constructing and studying analogous systems based on other physical phenomena. Prominent among these are studies of frustrated arrays of superconducting loops, a line of work that was initiated in the mid 1990's (Davidovic 1996, 1997). They demonstrated clear evidence for local antiferromagnetic correlations, and further work studied both frustrated and unfrustrated arrays of Josephson junctions (Hilgenkamp, 2003). Both systems were affected by intrinsic structural disorder, although this was less of a factor in the Josephson junction arrays. Even in those systems, however, only short-range correlations were observed and there were no clear distinctions between frustrated and unfrustrated arrays.

Geometric frustration leading to spin ice behavior has been explored in macroscopic systems as well. Confined layers of colloidal particles naturally form a triangular lattice, and their offset above and below the average layer height creates an Ising-like system (with thermal fluctuations and strength of interactions tunable through density and temperature) (Han, 2008). In addition, Mellado and coworkers recently realized a macroscopic two dimensional honeycomb array of magnetic rotors (Mellado 2012) which obeys the pseudo ice rule. Such a macroscopic system possesses the advantage of allowing an explicit study of nonlinear dynamics, with emerging domain walls and novel solitons, although it does not allow for the thermalization and electrical transport studies that have recently emerged for the nanoscale systems.

In addition to these experimentally realized artificial spin ice systems, there have been theoretical proposals for spin ice analogs elsewhere, such as nanostructured superconductors and optical traps (Libal, 2006; Libal, 2009). As these variants offer additional flexibility in how ice-like systems can be probed and how fluctuations and disorder can be introduced, they hold great promise as complements to the magnetic systems.

## VII. Future prospects

The study of artificial spin ice is still in its early days, and there are a number of directions into which the field is likely to move in the coming years. The fabrication of these systems is becoming easier with advancing lithography techniques, and experience from the initial years of study provides a strong foundation on which to probe the physics more deeply.

From the experimental side, we expect a wider search for exotic physics in non-trivial lattices other than the square and honeycomb (Morrison 2013; Chern 2012) for in and out of plane spins (Zhang, 2012). Furthermore, the ability to realize a more docile and dynamical material, that responds to temperature while still being directly imaged, underlies the possibility of further thermodynamic studies, as well as of transport phenomena of magnetic charges.

One of the most important areas of artificial spin ice research in the near future will be to further probe the impacts of disorder in these systems. Disorder can take the form of variations in the nanoscale structure and surfaces of the islands and wires that form the constituent elements of these systems. While the abovementioned recent studies have probed the impact of such disorder both experimentally and theoretically, there remain enormous opportunities offered by controlling the nature and level of such disorder through lithography. Libal and coworkers have predicted that artificial spin ice systems should exhibit return point memory, which has implications for the study of disorder and domain wall pinning (Libal 2012). Disorder can also be introduced through local variations in the frustrated lattices–either altering individual islands to make them of different shape or simply making lattices with occasional islands missing. The physical phenomena associated with disorder in spin ice materials and other geometrically frustrated magnetic materials has been a topic of considerable interest for years (e.g. Villain 1979, 1980; Henley 1989; Schiffer 1997; Revell 2013), and the artificial spin ice systems offer new opportunities to fine tune the disorder and locally probe the consequences.

A host of new probes will also be applied to these systems, complementing the static imaging probes that have been used to date. In particular, significant advances are expected with electrical transport studies of connected networks – a technique that has already yielded important insights (Tanaka, 2006; Branford, 2012). The artificial spin ice field also has a natural connection to the emerging topics around magnonics, probing magnetic nanostructures with microwaves with frequencies that are tuned to correspond to magnetic excitations in the structures (Jain, 2010; Kruglyak, 2010). These new probes will be especially important as the field starts to examine new materials and systems in which the moments have thermal dynamics – the impact of which has already begun to appear (Kapaklis 2012; Morgan, 2011-2013; Branford 2012). Dynamic probes will be especially important as new techniques allow the systems to be scaled down in size to where room temperature thermal fluctuations may become important (Khajetoorians, 2012). As tiny AC currents of magnetic monopole excitations have been claimed to be realized experimentally in the pyrochlore materials (Giblin, 2011), one can easily imagine artificial spin ice to be realized as a medium for possible magnetricity by design.

The interplay of interactions and disorder has been a staple of condensed matter physics for decades, yet it is still among the less understood aspects of the field. The artificial spin ices and their microscopically observable real-time dynamics, in a setting where dynamical and quenched disorder feed off each other, places this family of materials snugly in between glasses and granular materials on one side, and the real-time thermally-excited dynamics of cold atomic systems on the other.

### Acknowledgements:

The authors are grateful to our collaborators and all the members of the artificial spin ice community for many insightful discussions and especially to Ian Gilbert and Gunnar Möller for careful reading of the manuscript. Figures are reproduced with permission from the authors, for which we thank them. Cristiano Nisoli was supported by the US Department of Energy at LANL under contract no. DEAC52-06NA253962, LDRD grant no. 20120516ER. Roderich Moessner especially thanks Gunnar Möller for collaborations on artificial spin ice. Peter Schiffer has been supported by the U.S. Department of Energy, Office of Basic Energy Sciences, Materials Sciences and Engineering Division under Grant No. DE-SC0005313.

Note added: In this fast-moving field, new developments continue to arrive in rapid succession, and we apologize to authors of recent work that was not available at the time of the preparation of this manuscript.